\documentclass[12pt]{article}
\usepackage{amsfonts,a4}
\usepackage{pictex}
\usepackage{graphics}
\usepackage{amsmath}
\usepackage{amssymb}
\newcommand{\nc}{\newcommand}

\nc{\di}{\displaystyle}
\nc{\nn}{\nonumber}
\nc{\nek}{\nonumber\\[1ex]}
\nc{\st}{\scriptstyle}
\nc{\tst}{\textstyle}
\nc{\nb}{\normalsize\bf}
\nc{\ns}{\normalsize}
\nc{\seq}{\subseteq}
\nc{\AH}{\mathcal{A}}
\nc{\BE}{\mathcal{B}}
\nc{\CE}{\mathcal{C}}
\nc{\D}{\mathcal{D}}
\nc{\E}{\mathcal{E}}
\nc{\EF}{\mathcal{F}}
\nc{\GE}{\mathcal{G}}
\nc{\HA}{\mathcal{H}}
\nc{\J}{\mathcal{J}}
\nc{\KA}{\mathcal{K}}
\nc{\EL}{\mathcal{L}}
\nc{\PE}{\mathcal{P}}
\nc{\ER}{\mathcal{R}}
\nc{\ES}{\mathcal{S}}
\nc{\TE}{\mathcal{T}}
\nc{\EM}{\mathcal{M}}
\nc{\EN}{\mathcal{N}}
\nc{\OH}{\mathcal{O}}
\nc{\U}{\mathcal{U}}
\nc{\WE}{\mathcal{W}}
\nc{\EX}{\mathcal{X}}
\nc{\Y}{\mathcal{Y}}
\nc{\ZE}{\mathcal{Z}}
\nc{\ma}[1]{\mbox{$\,{#1}\,$}}
\nc{\ek}{\protect\\[1ex]}
\nc{\zx}{\protect\\[2ex]}

\newcommand{\R}{{\mathbb R}}

\nc{\IR}{\mbox{\bf R}}
\nc{\IN}{\mbox{\bf N}}
\nc{\ZZ}{\mbox{\bf Z}}
\nc{\la}{\lambda}
\nc{\La}{\Lambda}
\nc{\da}{\delta}
\nc{\Da}{\Delta}
\nc{\ta}{\theta}
\nc{\Ta}{\Theta}
\nc{\na}{\nabla}
\nc{\ue}{\infty}
\nc{\vp}{\varphi}
\nc{\vta}{\vartheta}
\nc{\Gm}{\Gamma}
\nc{\gm}{\gamma}
\nc{\ka}{\kappa}
\nc{\si}{\sigma}
\nc{\Si}{\Sigma}
\nc{\al}{\alpha}
\nc{\be}{\beta}
\nc{\om}{\omega}
\nc{\Om}{\Omega}
\nc{\pa}{\partial}
\nc{\ti}{\times}
\nc{\n}{|}
\nc{\rub}{\,\rule[-2.7pt]{.02in}{4mm}\,}
\nc{\ab}{\|}
\nc{\s}{\tilde}
\nc{\ve}{\varepsilon}
\nc{\fa}{\forall}
\nc{\ov}{\overline}
\nc{\un}{\underline}
\nc{\Llr}{\Longleftrightarrow}
\nc{\llr}{\longleftrightarrow}
\nc{\ra}{\rightarrow}
\nc{\lra}{\longrightarrow}
\nc{\rh}{\rightharpoonup}
\nc{\Ra}{\Rightarrow}
\nc{\ran}{\rangle}
\nc{\lan}{\langle}
\nc{\bs}{\backslash}
\nc{\ko}{\,,\,}
\nc{\eq}[1]{\mbox{\rm {(\ref{E#1})}}}
\nc{\qed}{\mbox{ }\nolinebreak\hfill \rule{2mm}{2mm}}
\nc{\ha}{\frac{1}{2}}
\nc{\kla}{\,[\,}
\nc{\klz}{\,]\,}
\nc{\lk}{\left[}
\nc{\rk}{\right]}
\nc{\lb}{\left\{}
\nc{\rb}{\right\}}
\nc{\rr}{\right)}
\nc{\lr}{\left(}
\nc{\f}{\big(}
\nc{\g}{\big)}
\nc{\Ba}{\Big(}
\nc{\Bz}{\Big)}
\nc{\Bka}{\Big[}
\nc{\Bkz}{\Big]}
\nc{\bka}{\big[}
\nc{\bkz}{\big]}
\nc{\Blb}{\Big\{}
\nc{\Brb}{\Big\}}
\nc{\blb}{\big\{}
\nc{\brb}{\big\}}
\nc{\pn}{\par\noindent}
\nc{\emp}{\emptyset}
\nc{\Ri}{\Rightarrow}
\nc{\hph}{\hphantom}
\nc{\vph}{\vphantom}
\nc{\vpn}{\vspace{2ex}\par\noindent}
\nc{\vpar}{\vspace{2ex}\par}
\nc{\mathe}[1]{\mbox{${\di {#1}}$}}
\nc{\equno}[1]{\\[-.5ex] \mbox{}\label{#1}\\[-.5ex]}
\nc{\tr}{{\mathrm{tr}}\,}
\newcommand{\Cm}{C_m^{(0)}}
\begin{document}
\title{{\Large {\bf Materials with a desired refraction coefficient can be made by embedding small particles.
}}}
\author{A. G. Ramm \\ {\ns (Mathematics Department, Kansas St. University,}\\ 
{\ns Manhattan, KS66506, USA} \\ {\ns and TU Darmstadt, Germany)}\\ 
{\small ramm@math.ksu.edu}}
\date{}
\maketitle
\begin{abstract}\noindent
A method is proposed to create materials with a desired refraction 
coefficient, possibly negative one. The method consists of embedding 
into a given material small particles. Given $n_0(x)$, the refraction 
coefficient of the original material in a bounded domain 
$D \subset \R^3$, and a desired refraction coefficient $n(x)$, one 
calculates the number $N(x)$ of small particles, to be embedded in $D$ 
around a point $x \in D$ per unit volume of $D$, in order that the 
resulting new material has refraction coefficient $n(x)$.
\end{abstract}

{\small PACS: 03.04.Kf\\ MSC: 35J05, 35J10, 70F10, 74J25, 81U40, 81V05} 
\ek {\small Keywords: "smart" materials, wave scattering by small bodies, 
many-body scattering problem, negative refraction, nanotechnology} 

\section{Introduction }\label{S1} 
There is a growing interest to materials with the desired 
properties, in particular, with negative refraction coefficient (see 
\cite{A} and references therein). In \cite{AG} the role of spatial 
dispersions is emphasized in explaining unusual properties of materials. 
In \cite{L} the role of dispersion for wave propagation in
solids is described.
In \cite{MK} boundary-value problems in domains with complicated 
boundaries were studied. In \cite{R476}, \cite{R524} wave scattering by 
small bodies of arbitrary shapes is studied and formulas for the 
$S$-matrix are obtained. In \cite{R515} a general method for creating 
materials with wave-focusing properties is proposed and justified. Our aim 
in this paper is to use a similar approach for creation of the materials 
with a desired refraction coefficient by embedding small particles into a 
given material with known refraction coefficient $n_0(x)$. The acoustic 
wave scattering by the given material is described by the Helmholtz 
equation \begin{equation}
	\label{E1}
	[\na^2 + k^2 n_0(x)] u=0 \text{ in } \R^3, \quad n_0(x) = \begin{cases}
	1 & \text{ in } D':=\R^3\setminus D,\\
	n_0(x) \quad & \text{in } D.
	\end{cases}
\end{equation}
Here $k>0$ is the wavenumber in $D'$. Equation \eq{1} can be 
written as the Schr\"odinger equation
\begin{equation}
	\label{E2}
	L_0 u :=[\na^2 + k^2-q_0(x)] u=0 \text{ in } \R^3, 
\quad q_0:= k^2-k^2 n_0(x).
\end{equation}
We assume $k>0$ fixed and do not show $k$-variable in $q_0$. 
Clearly, $q_0=0$ in $D'$. The scattering solution to \eq{2} is 
uniquely defined by the radiation condition:
\begin{equation}
\label{E3}
u_0 = e^{ik \alpha \cdot x}	+ A_0(\be,\al)\, \frac{e^{ik r}}{r} +
o(\frac{1}{r}),\quad r := \n x\n \ra \ue,\: \be:= \frac{x}{r}\,.
\end{equation}
Here $\al \in S^2$ is a given unit vector: the direction of the incident 
plane wave, $S^2$ is the unit sphere in
$\R^3$, $A_0(\be,\al)$ is the scattering amplitude, and $\be$ is the
unit vector in the direction of the scattered wave.

Assume that $M$ small particles $D_m$, $1\leq m\leq M$, are embedded
into $D$. Smallness means $n_0 ka \ll 1$,  where $a = 0,5 \max_m \, {\rm
diam}\, D_m$, and $n_0= \max_{x\in D}\n n_0(x)\n$. On the boundary $S_m$
of $D_m$ an impedance boundary condition is satisfied: 
$$u_N(s) = \zeta_m u(s), \quad s\in S_m, \quad 1\leq m \leq M, $$
where $N$ is the unit normal to $S_m$ pointing out of $D_m$.
We assume that the surface $S_m$ is Lipschitz, and the Lipschitz constant 
does not depend on $m$, $1\leq m \leq M$.
The scattering problem can now be stated as follows:
\begin{equation}
\label{E4}
L_0 u = 0\quad \mbox{in } \R^3\bs \bigcup^M_{m=1}\, D_m,\quad u_N =
\zeta_m u\quad \mbox{on }S_m,\quad 1\leq m\leq M,
\end{equation}
\begin{equation}
\label{E5}
u(x) = u_0(x) + A_M(\be,\al)\, \frac{e^{ikr}}{r} +
o\big (\frac{1}{r}\big),\quad r =\n x\n\ra \ue,\quad \be = \frac{x}{r}\,.
\end{equation}
We prove that the solution to problem \eq{4} -- \eq{5} converges as
$M\ra \ue$ to the solution of the problem
\begin{eqnarray}
\label{E6}
L \U &:= & [\na ^2 + k^2-q(x)]\U = 0\quad \mbox{in }\R^3,\\
\U & = & e^{ik\al\cdot x} + A(\be,\al)\, \frac{e^{ikr}}{r} +
o\big(\frac{1}{r}\big),\quad r = \n x\n \ra \ue,\: \be = \frac{x}{r}\,,\label{E7}
\end{eqnarray}
where
\begin{equation}
\label{E8}
q(x)=q_0(x) + p(x),
\end{equation}
and give a formula for $p(x)$. It turns out that $p(x)$ can be made an
arbitrary desired function by choosing the density of the number $N(x)$
of the embedded particles  around each point $x\in D$ and the impedances
$\zeta_m$ properly. Thus, $q(x)$ can be made an arbitrary desired
function. Therefore the refraction coefficient
\begin{equation}
\label{E9}
n(x)=1-k^{-2}q(x)=n_0(x)-k^{-2}p(x)
\end{equation}
can be made arbitrary, in particular, negative.

If $n_0(x)$ is given and one wishes to create the material with the
coefficient $n(x)$, then one calculates 
$$p(x) = [n_0(x)-n(x)]k^2,$$ 
and
embeds $N(x)$ small particles per unit volume of $D$ around each 
point
$x\in D$ and chooses their impedances $\zeta_m$ so that the 
function $p(x)$ is obtained for the new material. In Section \ref{S2} we give
analytical formulas for $N(x)$ and $\zeta_m$ and sufficient
conditions for the convergence of the solution to \eq{4} -- \eq{5} to
the solution of \eq{6} -- \eq{8} as $M\to \infty$ in such a way that 
relations \eq{13}-\eq{14} hold.

We also prove that the relative volume of the embedded particles
is negligible. More precisely, if  $\n D_m\n$ the volume of $D_m$, then
$$\lim_{M\to \infty}\frac{\sum^M_{m=1} \, \n D_m\n}{D}=0.$$ 

By $\n S_m\n$ we denote the surface area of
$S_m$. We use an approximate formula for the electric
capacitance of the perfect conductor with boundary $S$ (see
\cite{R476}, p.26, formula (3.12)):
\begin{equation}
\label{E10}
C_m^{(0)}\approx
 \frac{4\pi\n S_m\n^2}{J_m},\quad  J_m:= \int_{S_m}\int_{S_m} \frac{ds\, 
dt}{\n 
s-t\n}\,.
\end{equation}
Note that 
$$\Cm = O(a), \quad\n S_m\n = O(a^2), \quad J_m=O(a^3).$$
By $\Cm$ the electric capacitance of a perfect conductor with the surface 
$S_m$ is denoted. 
We assume that
\begin{equation}
\label{E11}
n_0ka\ll 1, \quad d \gg a,\quad d:= \min_{m\neq j} {\rm dist}(D_m,D_j).
\end{equation}
Let 
\begin{equation}
\label{E12}
C_{m\zeta_m}:= \Cm \bka 1 + \Cm (\zeta_m \n S_m\n)^{-1}\bkz^{-1}.
\end{equation}
We assume throughout the paper that 
$$ d = O(a^{1/3}), 
\,\,a= O\big(\frac{1}{M}\big).$$ 
Let $M\ra \ue$ and assume that the following limit 
exists:
\begin{equation}
\label{E13}
\lim_{\substack{M\ra \ue\\ \n
x_m-x\n\leq d}}\Cm (\zeta_m\n S_m\n)^{-1}:= h(x).
\end{equation}
Here and below $x_m \in D_m$ is an arbitrary point in $D_m$. Because 
$D_m$ is small, the choice of this point in $D_m$ is not important.
Under the assumed relations between $a$ and $d$ one has
$\lim_{M\ra \ue}\frac{a}{d} = 0.$ The limit \eq{13} exists if and only if
$\zeta_m=O(a^{-1})$, because $|S_m|=O(a^2)$ and $\Cm=O(a)$.

Denote by $N_m(x)$ the number of small particles per unit volume around
a point $x\in D$: $\int_{\s{D}}N_M(x)dx = \sum_{D_m\subset \s{D}} 1$ for
any subdomain $\s{D}\subset D$.

The number of particles per unit volume is $O\f \frac{1}{d^3}\g =
O\f\frac{1}{a}\g,$. Therefore  their relative volume is $O\f 
\frac{a^3}{d^3}\g =
O(a^2)\ra 0$ as $M\ra \ue$. On the other hand, the quantity
$N_M(x)C_{m\zeta_m}$, which has physical meaning of the average quantity
$C_{m\zeta_m}$ per unit volume of $D$ around point $x$, has a limit:
\begin{equation}
\label{E14}
\lim_{\substack{M\ra \ue\\ \n x_m-x\n\leq d}} N_M(x) C_{m\zeta_m} =
\frac{C(x)}{1 + h(x)}\,,\quad \lim_{\substack {M\ra \ue\\ \n x_m-x\n
\leq d}} N_M(x)C_m^{(0)} := C(x).
\end{equation}
The existence of the finite second limit in \eq{14} is clear because
$N_M(x) = O\f \frac{1}{a}\g$ and $C_M^{(0)}=O(a)$, and the existence of
the first limit in \eq{14} follows from formula  \eq{13}
and from the second formula  \eq{14}. Our basic result is the formula:
\begin{equation}
\label{E15}
\bka n_0(x)-n(x)\bkz k^2:= p(x)= \frac{C(x)}{1 + h(x)}\,,
\end{equation}
where $C(x)$ is defined in \eq{14} and $h(x)$ is defined in \eq{13}.
\vpn
{\bf Example 1.} Suppose $\zeta_m=\ue$, so $\U\n_{\textstyle S_m}=0$,
which corresponds to acoustically soft particles. Then $h(x)=0$, 
$p(x)=C(x)$.
Assume that the small particles are balls of radius $a$. Then $\Cm = 
a$,
$N_M(x) = \frac{p(x)}{a}\,$, $M = O\f \frac{1}{a}\g$. Since $N_M(x) >
0$ and $C^{(0)}_m>0$, then $p(x)\geq 0$, so one can create in this case 
only non-negative functions $p(x)$. For any
positive function $p(x)$ one should embed $N(x) = \frac{p(x)}{a}$ small
acoustically soft balls of radius $a$ per unit volume of $D$ around each 
point
$x\in D$, and the resulting material will have $n(x) = n_0(x)-k^{-2}p(x)$. 
In
particular, $n(x) < 0$ if $p(x) > k^2 n_0(x)$.
\vpn
{\bf Example 2.} Choose an arbitrary function $p(x)=p_1(x) + ip_2(x)$,
$p_2(x) \leq 0$. The condition $p_2 \leq 0$ guarantees uniqueness of the
solution to problem \eq{6}-\eq{7} with $q (x)=q_0(x) + p(x)$.
Physically this condition means that the medium, corresponding to $n(x) =
1-k^{-2}q(x)$ has nonnegative absorption. Let the particles be balls of
radius $a$ and $\zeta_m = \zeta_m(x) = \frac{1}{4\pi ah(x)}$, where $h(x)$
is an arbitrary function at the moment. This function is fixed later. Then 
formula \eq{13} holds
because $\n S_m\n = 4\pi a^2$. Choose $N=N(x)$ and $h(x) = h_1 + ih_2$
from the first equation \eq{14} using \eq{12}:
$$
p_1 + p_2 = \frac{Na}{1 + h(x)} = \frac{Na (1 + h_1 - ih_2)}{(1 + h_1)^2
+ h_2^2}\,. $$
Thus,
\begin{equation}
\label{E16}
p_1 = \frac{Na(1+h_1)}{(1 + h_1)^2 + h_2^2}\,,\quad p_2 = -\frac{Na
h_2}{(1 + h_1)^2 + h_2^2}\,.
\end{equation}
We have three functions: $N=N(x) > 0$, $h_1$ and $h_2$, to satisfy two
equations \eq{16}. This can be done by infinitely many ways. For
instance, one can take $h_1 = 0$,  $h_2 = - \frac{p_2}{p_1}\,$
and $N =a^{-1}\, p_1 \f 1 + \frac{p_2^2}{p_1^2}\g$. Thus, to get the material
with the desired $n(x) = n_0(x) - k^{-2}p(x)$, where $p(x) = p_1(x) + i 
p_2(x),$ one
embeds $N(x) = a^{-1}(p_1^2 + p_2^2)/p_1$ small balls of radius $a$ per
unit volume around each point $x$ and chooses the impedance $\zeta_m(x)
= \f 4\pi ah(x)\g^{-1}$, where $h = h_1 + ih_2$, $h_2 = - p_2/p_1$, $h_1
= 0$. 

\section{Derivation of the results.}\label{S2}
We seek the unique solution to \eq{4} -- \eq{5} of the form 
\begin{eqnarray}
\label{E17} u &= & u_0 + \sum^M_{m=1} \int_{S_m} G(x,t)\si_m(t)dt \\
\nn & = & u_0 + \sum^M_{m=1} G(x,x_m) Q_m + \sum^M_{m=1} \int_{S_m} \bka
G(x,t)-G(x,x_m)\bkz{\si_m \, dt}.
\end{eqnarray}
Here $L_0 G_1 = \da(x-y) $ in $\R^3$, $G$ satisfies the radiation
condition, $\si_m$ are to be chosen so that the boundary condition \eq{4}
is satisfied, $Q_m:= \int_{S_m} \si_m\, dt$, $x_m\in D_m$. In the
generic case $Q_m\neq 0 $ one can neglect the last term in \eq{17}
compared with the preceding term if $\n x-x_m\n > d \gg a$ for all
$m$. Indeed, under this assumption one has $\n G(x,t)-G(x,x_m)\n\leq \n
\na_y G(x,\s{y})\cdot (t-x_m)\n = O\f \frac{a}{d}\g \ll 1$, where
$\s{y}:=x_m+\tau(t-x_m), \,\, 0<\tau<1,$ is a 'middle point'.  Thus, the 
third
term on the right side of \eq{17} is $O\f \frac{a}{d}\, \n Q_m \n\g\ll \n
Q_m\n$, where we also assume that $\n Q_m\n = O\f \int_{S_m} \n \si_m\n
\, dt\g$. We will see that this assumption is justified. For
example, if $u|_{\textstyle {S_m}}=0$, then $\si_m$ does not change sign 
on $S_m$.
Thus, generically one can write
\begin{equation}
\label{E18}
u= u_0(x) + \sum^M_{m=1} G(x,x_m)Q_m,\quad \n x-x_m\n\geq d \gg a,
\end{equation}
with the error $O\f \frac{a}{d}\g$, The choice of $x_m\in D_m$  does not
matter because $a$ is small. One may assume that $D_m$ are convex and
take $x_m$ at the gravity center of $D_m$.

The functions $G(x,y)$ and $u_0(x)$ are known because $n_0(x) $ is known.
Let us derive an equation for finding $Q_m$. If $Q_m$ are found then the
scattering problem \eq{4} -- \eq{5} is solved by formula \eq{18} for
any $x$ away from an immediate neighborhood of the small particles. To
derive an equation for $Q_m$ we need some preparations. The function
$G(x,y)$ solves the equation:
\begin{equation}
\label{E19}
G(x,y)=g(x,y)-\int_D g(x,z)q(z)\, G(z,y)dz, \quad g(x,y):=
\frac{e^{ik\n x-y\n}}{4\pi \n x-y\n}\,.
\end{equation}
One can easily prove that
\begin{equation}
\label{E20}
G(x,y)=g(x,y) \bka 1 + O(\n x-y\n)\bkz = g_0(x,y)\bka 1 + O(\n
x-y\n)\bkz,\quad \n x-y\n \ra 0,
\end{equation}
where $g_0(x,y)= (4\pi \n x-y\n)^{-1}$. Let
\begin{equation}
\label{E21}
T_j\si_j:= \int_{S_j} G(s,t)\, \si_j(t) dt,\quad A_j\si_j = 2 \int_{S_j}
\frac{\pa g_0(s,t)}{\pa N_s}\, \si_j(t)dt.
\end{equation}
It is known (\cite{R476}, p. 91 ) that
\begin{equation}
\label{E22}
\int_{S_j}A_j \si_j\, dt = - \int_{S_j}\si_j(t)dt,\quad \frac{\pa
(T_j\si_j)}{\pa N_s} = \frac{A_j(k)\si_j-\si_j}{2}\,,
\end{equation}
where $A_j(k)$ is the operator similar to \eq{21} with $g(s,t)$ in place
of $g_0(s,t)$, $N_s:= N$ is the outer normal to $S_j$ at the point $s\in
S_j$. On the surface $S_j$ we have
\begin{equation}
\label{E23}
u=u_e(s) + T_j\si_j,\quad u_e:= u_0 + \sum^M_{m\neq j} G(s,x_m)Q_m.
\end{equation}
Using boundary condition \eq{4} and formulas \eq{22}, \eq{23}, one gets
\begin{equation}
\label{E24}
u_{e_N}(s)-\zeta_j u_e(s) + \frac{A_j\si_j-\si_j}{2} - \zeta_j T_j \si_j
= 0.
\end{equation}
Integrate \eq{24} over $S_j$, use \eq{22} and get:
\begin{equation}
\label{E25}
Q_j = \int_{S_j} u_{e_N} (s)ds - \zeta_j \int_{S_j} u_e(s)ds - \zeta_j
\int_{S_j} T_j \si_j\, ds.
\end{equation}
One has 
$$\int_{S_j} u_{e_N} ds = \int_{D_j} \Da u_e dx = O(k^2 
a^3),\quad
\int_{S_j} u_e ds =u_e(x_j)|S_j|= O(a^2),$$
where  the smallness of $D_j$ and the fact that $u_e$ and its 
two derivatives are bounded on
$S_j$ were used. Since $ka \ll 1$ we can neglect the first integral in 
\eq{25}
compared with the second. Furthermore 
$$I: = \int_{S_j} T_j \si_j ds =
\int_{S_j} dt \si_j(t)\int_{S_j} \frac{ds}{4\pi \n s-t\n}\,.$$ 
We replace
the last integral by its mean value 
$$\frac{1}{\n S_j\n} \int_{S_j}dt
\int_{S_j} \frac{ds}{4\pi \n s-t\n} := \frac{J_j}{4\pi \n S_j\n}\,.$$
Thus, $I = \frac{J_j Q_j}{4\pi \n S_j\n}\,$, and \eq{25} yields:
\begin{equation}
\label{E26}
Q_j =- \frac{\zeta_j\n S_j\n }{1 + \zeta_j J_j (4\pi \n S_j\n 
)^{-1}}  u_e(x_j)\,.
\end{equation}
We have replaced $u_e(s)$ by $u_e(x_j)$ because $\n x_j-s\n < 2a$ and $a$ 
is
small while $u_e(x)$ is continuous in a neighborhood of $x_j$. Using
\eq{10} we rewrite \eq{26} as
\begin{equation}
\label{E27}
Q_j = - C_j^{(0)} \bka 1 + C_j^{(0)} \f \zeta_j \n S_j\n)^{-1}]^{-1}\,
u_e(x_j):= -C_{j \zeta_j} \, u_e(x_j).
\end{equation}
Thus, \eq{18} can be written as:
\begin{equation}
\label{E28}
u(x) = u_0(x)-\sum^M_{m=1} G(x,x_m) C_{m \zeta_m}
 \, u(x_m) ,\quad \n x-x_m\n \geq d \gg a.
\end{equation}
We have replaced $u_e(x_m)$ by $u(x_m)$ under the sign of the sum in
\eq{28} because at the points $x$ which are away from small particles
one has $u_e(x) = u(x)$ with the error $O\f \frac{a}{d}\g$. 
 Formulas \eq{13} -- \eq{14} allow
one to pass to the limit $M\ra \ue$ in \eq{28} and get
\begin{equation}
\label{E29}
\U(x) = u_0(x)-\int_D G(x,y)\, p(y)\, \U(y) dy,
\end{equation}
where $p(x) $ is defined in \eq{15}. 
Applying  to \eq{29}
the operator $L_0$, defined in  \eq{2},  and using
the relation $L_0G(x,y)=-\delta(x-y)$ yields 
equation \eq{6} with $q$ defined in
\eq{8}. The radiation condition  for $\U$ is satisfied:
\begin{equation}
\label{E30}
A(\be,\al) = A_0(\be,\al) + A_1(\be, \al),
\end{equation}
where 
\begin{equation}
\label{E31}
A_1(\be,\al) = \lim_{M\ra \ue}A_M(\be,\al) = - \frac{1}{4\pi} \int_D
u_0(y,-\be)\, p(y)\, \U(y)\, dy.
\end{equation}
Here we have used a result from \cite{R470}:
\begin{equation}
\label{E32}
G(x,y) = \frac{e^{ik\n x\n}}{4\pi \n x\n}\, u_0(y,-\be) + o\f
\frac{1}{\n x\n}\g,\quad \n x\n \ra \ue,\: \frac{x}{\n x\n} = \be.
\end{equation}
In our derivations it was assumed that $\zeta_m \neq 0$. If $\zeta_m = 0
$ for all $m$, that is, the small particles  are acoustically hard, then
$Q_m = 0$ in the first order
with respect to $ka$. One can show that in this case $Q_m=O(k^2a^3)$, and 
that the last sum in \eq{17} is of
the same order of magnitude as the preceding sum. Consequently, 
the theory in this case is quite different: the effective field in the 
medium is not 
described by equation \eq{29}, which is equivalent to a local
equation 
\eq{6}. In fact, the effective field in this case is described by an 
integrodifferential equation which is not equivalent to a local 
differential equation.
  
Let us  
explain the relation  $Q_m=O(k^2a^3)$, mentioned above. Write \eq{24} with 
$\zeta_j 
= 0$, integrate over $S_j$ and use the first formula \eq{22} to get 
$$ Q_j = \int_{S_j} u_{e_N} ds = \int_{D_j} \Da u_e\, dx = O(k^2 a^3).$$

\end{document}